\documentclass[cameraready]{Interspeech}

\usepackage{algorithm}
\usepackage{algpseudocode}
\usepackage{amsmath,amssymb}
\usepackage{multirow}

\usepackage{caption}
\title{Imitation Learning for Elder-Facing Speech Synthesis}

\author[affiliation={1}]{Dongrui}{Han}
\author[affiliation={1}]{Weidong}{Chen}
\author[affiliation={1}]{Jiawen}{Kang}
\author[affiliation={2}]{Mingyu}{Cui}
\author[affiliation={1}]{Helen}{Meng}
\author[affiliation={1}, correspondingauthor]{Xixin}{Wu}

\address{
    $^1$ The Chinese University of Hong Kong, Hong Kong SAR, China \\
    $^2$ Tencent Hunyuan, China
}

\email{\{drhan,wdchen,jwkang,hmmeng,wuxx\}@se.cuhk.edu.hk,\\ericmycui@tencent.com}

\keywords{speech synthesis, imitation learning, reinforcement learning, on-policy learning, elder-facing, reward hacking}

\usepackage{comment}

\begin{document}

\maketitle
\vspace{-1.5\baselineskip}
\begin{abstract}
\vspace{-0.3\baselineskip}
Recent advances in text-to-speech (TTS) synthesis have achieved highly natural and expressive speech generation. However, these systems are designed for general adults and overlook older adults’ speech comprehension needs due to age-related sensory and cognitive decline. Prior work involves older adults by collecting preference feedback to tune model parameters. However, obtaining sufficient preference data is costly and difficult, as older adults quickly become fatigued during collection. In this paper, we propose a novel imitation learning (IL) framework to learn TTS models from expert demonstrations. We further improve Group Relative Policy Optimization (GRPO) with two-stage on-policy reward learning (OPRL) to mitigate reward hacking under limited supervision from expert demonstration. Experimental results show that GRPO w/ OPRL outperforms GRPO and supervised baselines in objective and subjective metrics. Audio samples are available at \url{https://dongru1.github.io/demo/im-efss/}
\end{abstract}
\vspace{-1.5\baselineskip}
\section{Introduction}
\vspace{-0.5\baselineskip}
Text-to-speech (TTS) has advanced rapidly in recent years, reaching highly natural speech generation in many settings~\cite{chenDrawSpeechExpressiveSpeech2025,duCosyVoice3Inthewild2025,sunF5RTTSImprovingFlowMatching2025, minixhofer2026ttsds}, enabling speech-based interaction in various daily scenarios, for example, voice assistants, train announcements, and grocery self-checkouts~\cite{herrmannPerceptionArtificialintelligenceAI2023a}. 
Meanwhile, human populations are rapidly ageing worldwide due to declining birth rates and rising life expectancies~\cite{khanPublicHealthChallenges2024}. Due to inevitable age-related sensory and cognitive deterioration, older adults' speech comprehension is different from that of younger adults, e.g., higher hearing thresholds~\cite{herrmannPerceptionArtificialintelligenceAI2023a,suttonYoungerOlderAdults1995,woltersMakingSpeechSynthesis2007,krauseSpeakingClearlyOlder2019,chenHILvoiceHumanintheLoopStyle2022}. However, adapting TTS systems to address these differences remains challenging~\cite{herrmannPerceptionArtificialintelligenceAI2023a}. 

One intuitive solution is to involve older adults in the development of TTS systems, by collecting their preference opinion on generated speech with various styles and tuning the TTS systems towards the style that is preferred by older adults. 
HILvoice~\cite{chenHILvoiceHumanintheLoopStyle2022} iteratively searches for controllable prosodic factors (e.g., duration and pitch) to generate an elder-preferred speaking style, updating these factors based on older adults' preference feedback. However, this human-in-the-loop optimization is time-consuming and costly because it relies on repeated subjective preference tests, and older adults may fatigue quickly during such tests. Consequently, it is difficult to scale this approach to large datasets.

To address this, we present an imitation learning framework that learns TTS models from expert demonstrations via inverse reinforcement learning (IRL)~\cite{ciosek2022imitation,10.5555/645529.657801,deshpandeAdvancesApplicationsInverse2025}. Instead of collecting preference data directly from older adults, we invite experienced healthcare professionals to record speech intended for older adults as expert demonstrations. The TTS models are trained to mimic the speaking style of the recorded speech by learning an expert reward model, which assigns high rewards to the expert-recorded speech and low rewards to other non-expert demonstrations (e.g., machine-generated speech), and finetuning TTS model parameters using the learned reward model. 

Recently, many works have applied reinforcement learning (RL) to improve TTS models~\cite{duCosyVoice3Inthewild2025,sunF5RTTSImprovingFlowMatching2025,zhang2024speechalign,anastassiouSeedTTSFamilyHighQuality2024,tianPreferenceAlignmentImproves2025,gaoEmoDPOControllableEmotional2025,chenFineTuningTexttoSpeechDiffusion2025a,zhangAdvancingZeroshotTexttospeech2025a,liuGroupRelativePolicy2025}. Liu et al.~\cite{liuGroupRelativePolicy2025} deploy Group Relative Policy Optimization (GRPO) to improve naturalness and expressivity of generated speech. Tian et al.~\cite{tianPreferenceAlignmentImproves2025} use generated pairwise data and deploy direct preference optimization (DPO) to train Language Model (LM)-based TTS models. Emo-DPO~\cite{gaoEmoDPOControllableEmotional2025} applies DPO on paired emotional speech generated by models to better capture nuanced emotion differences and enhance controllable emotional TTS. The rewards in these works are mainly based on objective metrics, e.g., word error rate (WER), as collecting human-labeled pairwise data is costly. Recent research also observes the reward hacking phenomenon~\cite{skalse2022defining,hu2025reward,shen2025exploring}, where the model focuses on maximizing the expected reward (towards a certain style preference) while ignoring other fundamental constraints (e.g., high speech quality and intelligibility) that are not specified in the reward function~\cite{anastassiouSeedTTSFamilyHighQuality2024,zhangAdvancingZeroshotTexttospeech2025a,duCosyVoice3Inthewild2025}.
Zhang et al.~\cite{zhangAdvancingZeroshotTexttospeech2025a} mitigate the reward hacking problem by constructing data pairs that reflect desired constraints, e.g., paired data of artificially distorted speech (with high WER) and unmodified data that are assigned with low and high rewards, respectively. However, such data construction significantly increases the training burden in order to impose sufficient constraints on the model. Moreover, as the TTS model is trained and gradually improved, the constraints that are not satisfied may change and the constructed data becomes useless.

In this paper, we propose to apply On-Policy Reward Learning (OPRL) to mitigate the reward hacking problem~\cite{lang-etal-2024-fine}. 
Specifically, we learn an expert reward model that explains what speech is preferred by older adults, from the expert demonstrations of healthcare professionals, and optimize the TTS model with GRPO under this expert reward model. 
Unlike previous RL-based TTS work, where the reward model is fixed while the TTS model is finetuned, we update the reward model gradually by adding GRPO rollouts (generated speech) to the reward model training dataset (i.e., with expert demonstrations and constructed negative samples) and retraining the reward model using the updated dataset. In this way, the comparison of the rollouts and the demonstrations provides a dynamic guidance for the TTS model training~\cite{ fuLearningRobustRewards2018}.
The key contributions of this work are threefold:
\begin{itemize}
    \item We propose a novel imitation learning framework for aligning TTS models using expert demonstrations.
    \item We propose an effective two-stage on-policy reward learning method to mitigate the reward hacking problem in RL-based TTS model training.
    \item We present an elder-facing TTS model that demonstrates effectiveness in generating elder-preferred speech, as verified by subjective listening tests conducted with older adults.
\end{itemize}
\vspace{-1\baselineskip}
\section{Methodology}

\begin{figure*}[t]
\vspace{-0.5\baselineskip}
    \centering
    \includegraphics[width=\textwidth]{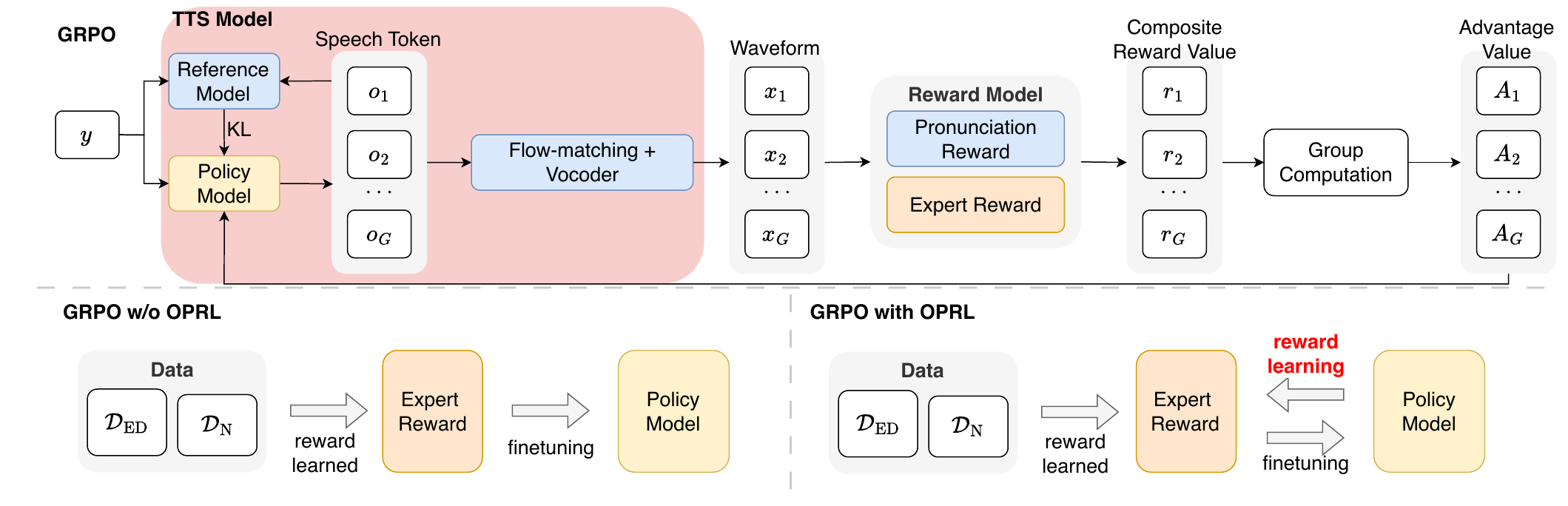}
    \caption{Overview of the GRPO pipeline and the on-policy reward learning (OPRL) process. 
    }
    \label{fig:oprl_pipeline}
\vspace{-0.5\baselineskip}
\end{figure*}
\vspace{-0.5\baselineskip}
\subsection{Imitation Learning}
\label{sec:il}
\vspace{-0.5\baselineskip}
Defining an appropriate reward function is challenging for speech synthesis targeting older adults, as quality is subjective and depends on multiple interacting factors (e.g., naturalness, intelligibility). Moreover, poorly specified rewards are prone to reward hacking, motivating the use of imitation learning to directly learn elder-facing speaking behaviors from expert demonstrations. Following Ciosek et al.~\cite{ciosek2022imitation}, we perform imitation learning with an RL algorithm by learning reward models that encourage generated speech similar to the expert demonstrations. Given expert demonstrations $\mathcal{D}_{\text{ED}}=\{y^{(i)}_w, x^{(i)}_w\}_{i=1}^N$ with recorded speech $x^{(i)}_w$ and the corresponding text $y^{(i)}_w$, negative samples $\mathcal{D}_{\text{N}}=\{y^{(i)}_l, x^{(i)}_l\}_{i=1}^N$ are either recorded using a neutral style or generated by a neutral-style TTS model. The reward model training set is denoted as $\mathcal{D}=\mathcal{D}_{\text{ED}}\cup\mathcal{D}_{\text{N}}$.
IL defines the rewards as
\begin{equation}
\mathcal{R}_{\text{int}}(x,y)=\mathbb{I}\big[(x,y)\in\mathcal{D}_{\text{ED}}\big],
\label{eq:ilr_indicator}
\end{equation}
and learns a reward model to guide a policy (i.e. TTS model) by maximizing the expected reward. The key idea is that high rewards are assigned to samples that align well with expert behaviors. Hence, the trained TTS model can imitate the expert behaviors. We introduce our improved multi-faceted rewards and OPRL in the following sections.

\vspace{-0.5\baselineskip}
\subsection{Multi-faceted Rewards}
\vspace{-0.5\baselineskip}
Our multi-faceted reward consists of expert reward and pronunciation reward, which separately guide the TTS model in terms of speaking style and intelligibility to alleviate the reward hacking problem.
\vspace{-0.5\baselineskip}
\subsubsection{Expert Reward}
\label{sec:elder_reward}
\vspace{-0.5\baselineskip}
The expert reward model, also referred to as intrinsic reward model~\cite{ciosek2022imitation}, aims to predict whether the generated speech is similar to expert demonstrations. 
The reward model consists of a frozen prosodic style encoder followed by a trainable reward head. 
We use the prosodic style encoder of a pretrained StyleTTS 2 model~\cite{liStyleTTS2Humanlevel2023} to extract prosodic embeddings from speech mel-spectrograms. 
The style encoder comprises a convolutional layer with spectral norm, $4$ residual blocks with spectral norm, spatial down-sampling, adaptive average pooling, and a linear projection, producing a $128$-dimensional prosodic embeddings. The encoder parameters are frozen during expert reward model training. The reward head applies LayerNorm~\cite{baLayerNormalization2016} to the $128$-dim embedding, then passes it through a stack of $6$ ResNet blocks~\cite{heDeepResidualLearning2016}. Each block contains $2$ linear layers with LayerNorm, ReLU activation, residual connection, and dropout ($p=0.1$). A final linear layer maps the representation to an unbounded scalar reward score $\hat{r}(x)$, which is normalized to $(0,1)$ via $\mathcal{R}_{\text{expert}}(x) = \frac{\tanh(\hat{r}(x)/2) + 1}{2}$.

We train the reward model with the Bradley--Terry pairwise ranking loss~\cite{bradleyRankAnalysisIncomplete1952}:
\begin{equation}
  \begin{aligned}
  \mathcal{L}_{\text{BT}}
  = - \mathbb{E}_{(x^{(w)}, x^{(l)})}
  \left[ \log \sigma \left( \hat{r}(x_w) - \hat{r}(x_l) \right) \right]
  \label{eq:bt-loss}
  \end{aligned}
\end{equation}
\noindent where $x_w$ denotes the preferred recorded speech utterance, $x_l$ the undesirable generated ones, and $\sigma(\cdot)$ the sigmoid function.

\vspace{-0.5\baselineskip}
\subsubsection{Pronunciation Reward}
\label{sec:pron_reward}
We use a pretrained Cantonese automatic speech recognition (ASR) model to transcribe speech into text.
Considering that there are lots of homophones in Cantonese, instead of using character error rate (CER), we adopt a syllable error rate (SER) based on Jyutping~\cite{tang2002guide} to compute the pronunciation reward. 
The raw SER $\in [0, +\infty)$ is then transformed into a bounded reward as $\mathcal{R}_{\text{pron}}(x,y) = \text{clip}(1 - \tanh(s_w \cdot \text{SER}(x,y)), 0, 1)$, with $s_w=3$ controlling the function steepness.
\vspace{-0.5\baselineskip}
\subsubsection{Composite Reward}
The two normalized rewards are combined via a weighted harmonic mean, analogous to the $F$-score:
\begin{equation}
  \begin{aligned}
  \mathcal{R}_{\text{comp}}(x,y) = \frac{2\mathcal{R}_{\text{Pron}}(x,y)\mathcal{R}_{\text{expert}}(x)}{\mathcal{R}_{\text{Pron}}(x,y)+\mathcal{R}_{\text{expert}}(x)}
  \label{eq:composite}
  \end{aligned}
\end{equation} 
The harmonic mean, unlike the arithmetic mean, aggressively penalizes low individual rewards, thus preventing the model from trading off intelligibility for prosody (or vice versa).
\vspace{-0.5\baselineskip}
\subsection{GRPO}
\vspace{-0.5\baselineskip}
In this work, we follow the GPRO for finetuning TTS models in~\cite{liuGroupRelativePolicy2025}, but keep the Proximal Policy Optimization (PPO) clip as in~\cite{shaoDeepSeekMathPushingLimits2024}. The overview of the pipeline is shown at the top of Figure~\ref{fig:oprl_pipeline}. Given a pretrained policy model $\pi_{\theta}(x|y)$, i.e. a TTS model, a group of speech token outputs $\{o_1,...,o_G\}$, a.k.a. rollout data, are generated based on an input text $y$. The tokens are converted into speech waveforms and fed to the trained composite reward model to obtain output rewards $R_g=\{r_1,...,r_G\}$. Group-relative advantages are estimated as $A_i=\frac{r_i-\text{mean}(R_g)}{\text{std}(R_g)}$.
Finally, the policy model is optimized by maximizing the following objective w.r.t the model parameters $\theta$: 
\begin{equation}
    \begin{aligned}
        \mathcal{J}_{GRPO}(\theta)
        =
        \mathbb{E}_{y\sim\mathcal{D}_{\text{ED}},\{{x}_i\}_{i=1}^{G}\sim \pi_{\theta}(x\mid y)}
        \\\frac{1}{G}
        \sum_{i=1}^{G}
        \frac{1}{\lvert o_i\rvert}\sum_{t=1}^{\lvert o_i\rvert}
        \min\Big(\rho_{i,t}A_i,\mathrm{clip}(\rho_{i,t},1-\epsilon,1+\epsilon)A_i\Big)\\
        \quad -\beta\, \mathbb{D}_{\mathrm{KL}}(\pi_{\theta}\,\Vert\,\pi_{\mathrm{ref}}),
    \end{aligned}
\end{equation}
where $\rho_{i,t} = \frac{\pi_{\theta}(o_{i,t}\mid\mathbf{y},o_{i,<t})}{\pi_{\theta_{\mathrm{old}}}(o_{i,t}\mid \mathbf{y},o_{i,<t})}$ is the importance sampling ratio with t denoting the token index. $\epsilon=0.2$ is the clipping range, and $|o_i|$ denotes the token number of $o_i$. $\beta$ is the coefficient of the $\mathrm{KL}$ penalty:
\begin{equation}
    \begin{aligned}
    \mathbb{D}_{\mathrm{KL}}(\pi_{\theta}\Vert\pi_{\mathrm{ref}})
    =
    \frac{\pi_{\mathrm{ref}}(o_{i,t}\mid \mathbf{y},o_{i,<t})}
    {\pi_{\theta}(o_{i,t}\mid \mathbf{y},o_{i,<t})}
    \\
    -\log(
    \frac{\pi_{\mathrm{ref}}(o_{i,t}\mid \mathbf{y},o_{i,<t})}
    {\pi_{\theta}(o_{i,t}\mid \mathbf{y},o_{i,<t})})
    -1,
    \end{aligned}
\end{equation}
where $\pi_{\text{ref}}$ is a frozen reference model initialized from the same weights as $\pi_{\theta}$ at the beginning of training. 
The KL penalty discourages $\pi_{\theta}$ from drifting too far from this reference model, improving training stability.
\vspace{-0.5\baselineskip}
\subsection{On-Policy Reward Learning}
\label{sec:fd}
Based on the GRPO pipeline, OPRL is a two-stage process that aims to iteratively update the reward model with the generated speech of the finetuned TTS model. 

\vspace{-0.5\baselineskip}
\subsubsection{Stage 1}
\label{sec:stage1}
\vspace{-0.5\baselineskip}

We initialize the policy $\pi_\theta$ by supervised fine-tuning on the expert demonstration dataset $\mathcal{D}_{\text{ED}}$. 
An initial expert reward model $\mathcal{R}_{\text{expert}}$ is trained on the reward model training set $\mathcal{D}$ using the Bradley--Terry loss in Eq.~(\ref{eq:bt-loss}). 
We perform an iterative process consisting of the following steps:
(1) following the standard GRPO pipeline to finetune the policy using $\mathcal{R}_{\text{expert}}$ based on the dataset $\mathcal{D}$, with the rollout data of all finetuning steps recorded; (2) selecting the rollout samples whose predicted expert rewards are within the 10th--90th percentiles and SER values are below $0.15$ to obtain $\mathcal{D}_{\text{gen}}$; (3) assigning the selected samples with a median reward that is higher than the negative samples and lower than the expert demonstrations in $\mathcal{D}$, and adding the samples to the dataset $\mathcal{D}=\mathcal{D}\cup\mathcal{D}_{\text{gen}}$, (4) finetuning the expert reward model $\mathcal{R}_{\text{expert}}$ with the updated dataset $\mathcal{D}$. This process is repeated by a predefined number $K$ set as 5 in this work.

\vspace{-0.5\baselineskip}
\subsubsection{Stage 2}
\label{sec:stage2}
\vspace{-0.5\baselineskip}
In stage 2, we aim to expose the reward model to speech data of more text variation in addition to the expert demonstration data. 
Although we can collect generated speech from the TTS model conditioned on external text data for reward model updates, the absence of corresponding demonstrations prevents direct fine-tuning of the reward model. In this regard, we assign reward values to the rollout samples based on their rankings. Based on the reward training data $\mathcal{D}$, the reward model $\mathcal{R}_{\text{expert}}$, the policy $\pi_\theta$ from Stage 1, and an additional text dataset $\mathcal{D}_{\text{txt}}$, Stage 2 is composed of the following steps: (1) finetuning the policy with GRPO based on $\mathcal{R}_{\text{expert}}$ using $\mathcal{D}_{\text{txt}}$ and recording the rollout data; (2) binning the rollout samples with $\text{SER}\leq 0.1$ into groups based on their SERs, with a bin size of 0.04; (3) for the bins with sufficient samples, ranking the samples based on predicted expert rewards, and assigning monotonically increasing rewards to samples at the 0th, 25th, 50th, 75th, and 100th percentiles to obtain the training data $\mathcal{D}_{\text{rank}}$; (4) adding the new data to the original training data $\mathcal{D}=\mathcal{D}\cup\mathcal{D}_{\text{rank}}$ and finetuning the expert reward model $\mathcal{R}_{\text{expert}}$ with the updated dataset $\mathcal{D}$; (5) the policy is finetuned again with GRPO using the latest expert reward model $\mathcal{R}_{\text{expert}}$.

\vspace{-1\baselineskip}
\section{Experiments}
\vspace{-0.5\baselineskip}
\begin{table*}[t]
\centering
\caption{Experimental results of elder-facing speech synthesis systems.
}
\label{tab:evaluation}
\resizebox{\textwidth}{!}{%
\begin{tabular}{l cc cccccccc c c}
\toprule
\multirow{2}{*}{\textbf{Model}} & \multicolumn{11}{c}{\textbf{Objective}} & \textbf{Subjective} \\
\cmidrule(lr){2-12} \cmidrule(lr){13-13}
 & SER$\downarrow$ & CER$\downarrow$ & PA$\uparrow$ & PE$\downarrow$ & EE$\downarrow$ & MCD$\downarrow$ & F0 Corr$\uparrow$ & F0 VR$\rightarrow$1.0 & $\text{Dur}_\text{sil}$(s) & Dur(s) & SIM$\uparrow$ & MOS$\uparrow$ \\
\midrule
Ground Truth & 13.79 & 9.08 & 100.0 & 0.00 & 0.00 & 0.00 & 1.00 & 1.00 & 5.43$\pm$0.63 & 19.27$\pm$2.15 & 1.00 & 3.45$\pm$0.18 \\
\midrule
CosyVoice2-Yue (base) & 14.89 & 7.52 & 38.66$\pm$3.55 & 2.04$\pm$0.25 & 2.68$\pm$0.23 & 7.37$\pm$0.71 & 0.64$\pm$0.06 & 0.94$\pm$0.07 & 3.39$\pm$0.66 & 14.52$\pm$1.41 & 0.65$\pm$0.04 & 2.53$\pm$0.24 \\
+ SFT & 9.93 & 4.91 & 53.33$\pm$4.12 & 1.27$\pm$0.21 & 2.24$\pm$0.27 & 4.87$\pm$0.60 & 0.83$\pm$0.04 & \textbf{1.02$\pm$0.05} & 5.42$\pm$0.60 & 19.62$\pm$1.71 & 0.77$\pm$0.02 & 3.55$\pm$0.14 \\
\ \ + GRPO w/o OPRL & 11.58 & 6.99 & 49.62$\pm$3.78 & 1.58$\pm$0.26 & 3.96$\pm$0.49 & 5.10$\pm$0.65 & 0.75$\pm$0.06 & 1.18$\pm$0.06 & 11.51$\pm$1.38 & 27.62$\pm$2.92 & 0.76$\pm$0.03 & 2.70$\pm$0.23 \\
\ \ + GRPO w/ OPRL Stage~1 & 8.27 & 4.38 & \textbf{57.70$\pm$4.15} & \textbf{1.16$\pm$0.26} & 2.20$\pm$0.22 & \textbf{4.77$\pm$0.55} & 0.81$\pm$0.06 & \textbf{1.02$\pm$0.05} & 5.95$\pm$0.51 & 20.60$\pm$1.71 & 0.77$\pm$0.02 & 3.54$\pm$0.14 \\
\ \ \ \ + GRPO w/ OPRL Stage~2 & \textbf{7.54} & \textbf{3.86} & 54.62$\pm$3.48 & 1.17$\pm$0.18 & \textbf{2.06$\pm$0.21} & 4.80$\pm$0.62 & \textbf{0.85$\pm$0.04} & 1.06$\pm$0.06 & 5.36$\pm$0.53 & 19.62$\pm$1.88 & \textbf{0.77$\pm$0.02} & \textbf{3.78$\pm$0.16} \\

\bottomrule
\end{tabular}%
}
\vspace{-0.5cm}
\end{table*}
\subsection{Experimental Setup}
\vspace{-0.5\baselineskip}

\textbf{Datasets.} 
We use an internal expert demonstration dataset comprising $125$ pairs of speech samples, with a total duration of $1.5$ hours. Each pair consists of two speech samples of the same content recorded by healthcare professionals from a local hospital: one sample is intended for older adults, and the other is in a neutral news broadcast style.
Drawing on their extensive experience, our goal is to enable the TTS model to learn from expert demonstrations from the professionals. 
The dataset is divided into three sets: train ($89$), dev ($18$), and test ($18$). 
For Stage 2, we use the ZoengJyutGaai dataset\footnote{\url{https://huggingface.co/datasets/CanCLID/zoengjyutgaai}} as the additional text data. We filter the data by retaining only text sentences longer than 5 characters to enhance prosodic differences. 
We only randomly select $5,000$ sentences for GRPO to make sure that each sentence can be used multiple times in a limited training time. 
With such a setup, we aim to increase the number of bins that contain a sufficient number of samples in stage 2.

\noindent\textbf{Models.}
We use pretrained Cantonese CosyVoice~2-Yue~\cite{duCosyVoice2Scalable2024,liWenetSpeechYueLargescaleCantonese2025} as the backbone, and perform supervised fune-tuning (SFT) on the expert demonstration data. The SFT model is then used as the initialization of the policy model. The CosyVoice~2-Yue is finetuned on WenetSpeech-yue dataset\cite{liWenetSpeechYueLargescaleCantonese2025}. For the pronunciation reward model, we use SenseVoice-small~\cite{anFunAudioLLMVoiceUnderstanding2024} as the ASR model, which is a multilingual ASR model trained with a dataset that includes $5,000$ hours of Cantonese speech data. \textit{GRPO w/o OPRL} is the model trained using GRPO without OPRL. \textit{GRPO w/ OPRL Stage 1} and \textit{GRPO w/ OPRL Stage 2} denote the final policy model from each stage of OPRL.

\noindent\textbf{Training Detail.} CosyVoice2-Yue is SFT using Adam (lr $1{\times}10^{-5}$) for $200$ epochs on $4$ GPUs. 
For training of the expert reward model $\mathcal{R}_{\text{expert}}$, we use AdamW (lr $1{\times}10^{-4}$, wd $0.01$) using a batch size of $32$, and $200$ epochs with early stopping (patience $100$).
The GRPO pipeline adopts AdamW (lr $1{\times}10^{-6}$, weight decay $0.01$), with a grad clip of $1.0$, grad accumulation of $2$, and batch size of $2$ with bfloat16. For each prompt, we sample $G{=}4$ rollouts using top-$k$ ($k{=}25$, $\tau{=}1.2$). A PPO clip of $\epsilon{=}0.2$ is adopted, and a KL penalty with $\beta{=}0.1$ is applied. We use early stopping with a patience of 1,000 steps. If the moving-averaged reward does not improve for 1,000 consecutive steps, training will be stopped. The reward moving-average window size is 20.

\noindent \textbf{Evaluation.}
We conducted subjective evaluations based on mean opinion score (MOS) experiments, by inviting 8 older adults to listen to generated Cantonese speech samples and provide preference scores on a 5-point scale (1 for definitely dislike and 5 for definitely like). The MOS experiment consists of 10 utterances per system. The participants are native Cantonese speakers, aged 66-83 (mean 71.2). For objective evaluations, we assess intelligibility via SER and CER. Prosody is measured by comparing synthesized and ground-truth speech utterances after dynamic time warpping (DTW) alignment~\cite{sakoe2003dynamic}, using the metrics of Pitch Accuracy (PA), Pitch mean absolute Error (PE), Energy mean absolute Error (EE), Mean 
Cepstral Distortion (MCD)~\cite{407206,battenbergLocationRelativeAttentionMechanisms2020}, F0 Correlation (F0 Corr) on voiced frames, log-F0 on jointly-voiced frames, and F0 Variance Ratio (F0 VR). 
PA is defined as the proportion of voiced frames in the ground truth for which the synthesized F0 ($\text{F0}^{\text{syn}}$) is within $\pm 1/4$ tone (50 cents) of the ground-truth F0 ($\text{F0}^{\text{gt}}$)~\cite{salamonMelodyExtractionPolyphonic2012}.
F0 VR is defined as $\text{std}(\text{F0}^{\text{syn}})/\text{std}(\text{F0}^{\text{gt}})$ on voiced frames. We also report silence duration ($\text{Dur}_{\text{sil}}$), total duration ($\text{Dur}$), and speaker similarity (SIM) using CAM++ embeddings~\cite{wang23ha_interspeech}. 95\% confidence intervals are also provided.

\begin{figure}[t]
    \centering
    \includegraphics[width=\linewidth]{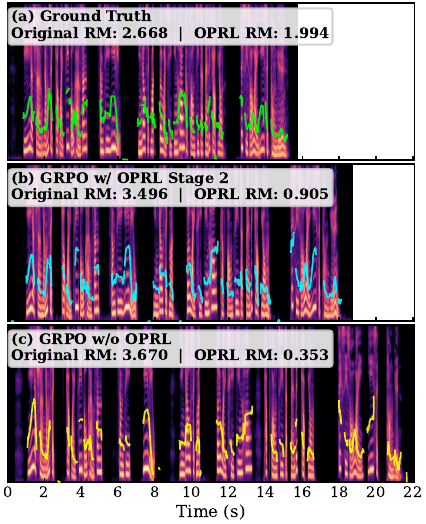}
    \caption{Generated mel-spectrograms and predicted rewards from original and OPRL-based reward models.}
    \label{fig:vis_mel}
\end{figure}
\vspace{-0.5\baselineskip}
\subsection{Results and Analysis}
\vspace{-0.5\baselineskip}

The evaluation results are summarized in Table~\ref{tab:evaluation}. GRPO w/o OPRL exhibits clear signs of suffering reward hacking. Its $\text{Dur}_\text{sil}$ value of 11.51 is more than double that of the ground truth, 5.43, and its Dur value of 27.62 exceeds the ground truth 19.27 by over 43$\%$. However, its SER (11.58$\%$) and CER (6.99$\%$) are inferior to the baseline SFT model's performance, and the other objective metrics are also weaker than those of SFT. The MOS of GRPO w/o OPRL is among the lowest, i.e., the generated speech is less preferred by older adults. These findings suggest that the model overemphasizes the slow speaking rate exhibited by the experts, inserting too many pauses and elongating the generated speech, rather than genuinely improving prosodic quality while maintaining overall speech quality.

GRPO w/ OPRL Stage~1 outperforms GRPO w/o OPRL in all objective metrics. Compared with GRPO w/o OPRL, the proposed method effectively mitigates reward hacking while guiding the policy to better align with expert demonstrations. GRPO w/ OPRL Stage~2 further enhances the performance, with MOS significantly improved. Stage 2 achieves the best (or second) performance in a majority of the metrics, which demonstrates our proposed framework enables the policy model to align better with the expert demonstrations (even exceed the demonstrations) and attain superior intelligibility simultaneously. 
With more unseen text data added to Stage 2, the policy is exposed to a broader prosodic distribution than the expert demonstrations and achieves better prosodic metrics. The pronunciation reward also enforces the model to align with ground-truth speech in terms of intelligibility, as indicated by the improvements on SER and CER. Moreover, Stage 2 has an F0 VR of 1.06, which is the second highest among all models. This indicates that the policy tends toward a slightly higher, yet more natural pitch range with higher MOS. Statistical significance is assessed using pairwise Wilcoxon signed-rank tests~\cite{woolsonWilcoxonSignedrankTest2005} with Holm--Bonferroni correction~\cite{holm1979simple}. 
GRPO w/ OPRL Stage 2 achieves a significantly higher MOS than CosyVoice2-Yue and GRPO w/o OPRL ($p<0.01$), and it is also higher than all other baselines with ($p<0.05$).

We also visualize the mel-spectrograms and the corresponding pitch contours of ground-truth and generated speech in Figure~\ref{fig:vis_mel} for comparison. It can be observed that GRPO w/o OPRL generates speech with longer durations and more silences, but with a significantly lower MOS, which indicates that it encountered reward hacking. 
Each speech utterance is rated with two reward models, Original RM and OPRL RM, which denote the reward model trained with expert demonstrations and the OPRL-based reward model, respectively. It can be found that Original RM predicts higher rewards of generated speech, at 3.496 and 3.67, than that of ground-truth speech, at 2.668. On the contrary, the OPRL RM predicts lower rewards for the two generated speeches, providing meaningful guidance to the policy models.

\vspace{-0.9\baselineskip}
\section{Conclusion}
\vspace{-0.5\baselineskip}

This paper presents a novel imitation learning framework for elder-facing speech synthesis based on expert demonstrations. We propose the on-policy reward learning (OPRL) strategy, which iteratively refines the reward model with generated speech, for the GRPO pipeline to alleviate the reward hacking problem. The developed elder-facing TTS system is verified as effective from objective metrics and subjective evaluations conducted with older adults.
Experimental results show that GRPO w/ OPRL achieves the best performance in intelligibility and prosody, with the best MOS among all compared models, confirming the benefit of OPRL for low-resource preference alignment. 
In future work, we plan to investigate older adults' preferences for the textual structure of generated speech and analyze preference differences between older adults and general adults by directly including both groups in the evaluation.

\section{Acknowledgement}
This work is partially supported by the National Natural Science Foundation of China (62306260) and the Centre for Perceptual and Interactive Intelligence, a CUHK-led InnoCentre under the InnoHK initiative of the Innovation and Technology Commission, of the Hong Kong Special Administrative Region Government

\section{Use of Generative AI Disclosure}
During the preparation of this manuscript, the authors used generative AI tools exclusively for the purpose of language editing and manuscript polishing to improve readability. These tools were not used to generate any core scientific ideas, experimental data, or technical contributions. All authors have thoroughly reviewed and approved the final version of the manuscript, and assume full responsibility for the integrity and entirety of its content.

\bibliographystyle{IEEEtran}
\bibliography{mybib}

\end{document}